\documentclass[prl,twocolumn,a4paper,groupedaddress,showpacs,floatfix,aps,10pt,longbibliography,nobibnotes]{revtex4-2}
\usepackage{graphicx,amsmath,amssymb,amsfonts,dsfont,subfigure,color}

\newcommand{\mc}[1]{\ensuremath{\mathcal{#1}}}

\usepackage{color}

\makeatletter
\newcommand{\fmarki}{*}
\newcommand{\fmarkii}{*\ensuremath{\dagger}}
\newcommand{\fmarkiii}{\ensuremath{\dagger\dagger}}

\def\@fnsymbol#1{{\ifcase#1\or \fmarki\or \fmarkii\or \fmarkiii \else\@ctrerr\fi}}
\makeatother
\begin{document}

\title{Fast Single-shot Imaging of Individual Ions via Homodyne Detection of Rydberg-Blockade-Induced Absorption}

\author{Jinjin Du${}^{1}$}
\thanks{These two authors contributed equally.}
\author{Thibault Vogt${}^{1,2}$}
\email{ttvogt@mail.sysu.edu.cn}
\author{Wenhui Li${}^{1,3}$}
\email{wenhui.li@nus.edu.sg}

\affiliation{Centre for Quantum Technologies, National University of Singapore, 3 Science Drive 2, Singapore 117543${}^1$}
\affiliation{School of Physics and Astronomy, Sun Yat-sen University, Zhuhai, 519082, China${}^2$}
\affiliation{Department of Physics, National University of Singapore, 117542, Singapore${}^3$}


%
%
%

\begin{abstract}
We introduce well-separated $^{87}$Rb$^+$ ions into an atomic ensemble by microwave ionization of Rydberg excitations and realize single-shot imaging of the individual ions with an exposure time of 1 $\mu$s. This imaging sensitivity is reached by using homodyne detection of ion-Rydberg-atom interaction induced absorption. We obtain an ion detection fidelity of (80 $\pm$ 5)\% from analyzing the absorption spots in acquired single-shot images. These \textit{in situ} images provide a direct visualization of the ion-Rydberg interaction blockade and reveal clear spatial correlations between Rydberg excitations. The capability of imaging individual ions in a single shot is of interest for investigating collisional dynamics in hybrid ion-atom systems and for exploring ions as a probe for measurements of quantum gases.
\end{abstract}

\maketitle


Cold hybrid ion-atom systems are promising for studying a wide range of fundamental processes in many-body physics~\cite{harter2014atomionexp, tomza2019hybridionatom, cote2016ultracold}. These systems have emerged relatively recently, after cooling and trapping of both ions and atoms have been separately demonstrated. Investigations of cold collisions and chemical reactions between ions and atoms, sympathetic cooling of ions, charge mobility, long-range molecular ions, and quantum simulation of solid-state physics systems are being actively pursued~\cite{dieterle2021transport,feldker2020buffer,schmidt2020opticaltraps,weckesser2021ionFeshbach,hirzler2022feshbachdimers}. Of importance, ions immersed in cold atom gases are interesting as a test bed for polaron physics in solid-state systems~\cite{astrakharchik2021ionic,yan2020bose,camargo2018rydbergpolaron}.

The study of ion-atom hybrid systems would benefit greatly from a real-time non-invasive imaging technique. An imaging technique often employed for ions held in Paul traps is fluorescence imaging~\cite{neuhauser1980localized}. Whereas it has the advantage of being state selective and yields high optical resolutions, however it can only apply to a few possible species that possess a cycling transition in the optical frequency range. Moreover, it is relatively slow compared to the collisional timescales observed in ion-atom mixtures~\cite{RN7792}. Ion microscopy is a promising tool that has also been demonstrated for imaging cold atoms and ions. It achieves unprecedented resolution in this field of research and allows for the detection of any species of ions~\cite{stecker2017high,veit2021pulsed}. A disadvantage of this method is that it is destructive. An alternative method for imaging impurity ions immersed in a cloud of cold atoms has been demonstrated recently utilizing ion-Rydberg-atom interaction induced absorption of a probe light under the condition of electromagnetically induced transparency involving Rydberg states (Rydberg EIT)~\cite{gross2020ion}. However, the sensitivity of detecting individual ions has not been reached yet.

In this letter, we demonstrate single-shot imaging of individual ions immersed in an atomic gas with an exposure time of 1 $\mu$s. This is a drastic advancement from the work in Ref.~\cite{gross2020ion}. The main improvement lies in homodyning the Rydberg EIT imaging probe light with a strong reference beam for overcoming the camera read noise~\cite{kadlecek2001nondestructive}. The signal-to-noise ratio (SNR) becomes sufficient for identifying an absorption spot around a single ion with a probability of (80 $\pm$ 5)\%. We extract the statistics of the absorption shadows' amplitudes and sizes and find that they agree well with theory. Finally, the positions of ions allow us to reconstruct the spatial distribution and correlation of Rydberg excitations that are created in the fully blockaded regime and subsequently ionized with a microwave field to produce separated ions. Our work constitutes a key step in realizing real-time imaging of ions embedded in atomic gases, which will be a unique tool in studying the dynamics of cold hybrid ion-atom systems.

As in Ref.~\cite{gross2020ion}, the imaging technique relies on the high sensitivity of an upper Rydberg EIT level $|n,\ell \rangle$ to the spatially dependent electric field generated by an ion, where $n$ and $\ell$ are the principal and the orbital quantum numbers, respectively~\cite{vogt2007electric,PhysRevLett.121.193401}. More precisely, the quadratic Stark shift of the Rydberg state is given by $\Delta E_S= - C_4(|n,\ell\rangle )/R^4$ (for $\ell\leq 5$), where $R$ is the distance from the ion and $C_4$ scales as $n^{7}$ and highly depends on $\ell$. Inside the blockade sphere of radius $R_b= [2 C_4 / (\hbar \gamma_{EIT})]^{1/4}$, which is centered at the ion, the shift exceeds half of the imaging EIT linewidth $\gamma_{EIT}$ and the scattering of the probe light is enhanced. In order to resolve an individual ion in a single shot, the number of excess photons scattered inside the blockade sphere needs to be large enough to overcome imaging noises, consisting mainly of read noise and shot noise. In Ref.~\cite{gross2020ion}, we had chosen $|r\rangle = |27G_{9/2}, m_J = 9/2\rangle$ to maximize the blockade radius $R_b$ and consequently the excess photons scattered by an ion. This however was not sufficient for enabling single-shot imaging of individual ions on the microsecond timescale. The main challenge comes from contradicting requirements for the probe power. On the one hand, a low imaging probe power is necessary to achieve high EIT transmission in the absence of ions by minimizing the scattering due to the long-range interaction of atoms excited in the $|r\rangle$ state~\cite{gorshkov:11}. On the other hand, a high probe photon flux is required to surpass noises.

In this work, we interfere the weak probe light with a strong reference light, which are respectively the $\sigma^+$ and the $\sigma^-$ components of an elliptically polarized beam. This increases the photon count received by each camera pixel so that read noise and other noises are overcome. The probe light is on resonance with the $|g\rangle = |5S_{1/2}, F = 2, m_F = 2\rangle \rightarrow |e\rangle = |5P_{3/2}, F = 3, m_F = 3\rangle $ transition and has an input field strength $\text{E}_{P_0}$ before entering the atomic cloud. The $\sigma^-$ reference field of strength $\text{E}_R$ is detuned from the allowed weak transition $|g\rangle \rightarrow |5P_{3/2}, F = 3, m_F = 1\rangle$ by $2 \pi \times 23$~MHz in the presence of a 12.8 G bias magnetic field, and hardly interacts with atoms when passing through the cloud. The intensity ratio $r_{R-P_0}=|\frac{\text{E}_R}{\text{E}_{P_0}}|^2$ is adjusted with a variable wave retarder~\cite{supp}. Right after the atomic cloud, the probe field, carrying the information about the presence of ions, can be expressed as $\text{E}_S=\text{E}_{P_0} \sqrt{T} \exp \left(i\phi \right)$, where $T$ and $\phi$ are the transmission of the probe intensity and the phase change of the probe field, respectively. Both fields then pass through a lens system of transmission $T_i$ and magnification $M$, and are made to interfere using an analyzer. The resulting total intensity $I$ on the camera is
\begin{equation}
I=\frac{1}{4}\epsilon_0 c \ \frac{T_{i}}{M^2} \left(|\text{E}_S|^2+|\text{E}_R|^2+2 |\text{E}_S \text{E}_R| \cos \phi \right),
\end{equation}
where the additional factor of 1/2 is due to the PBS in front of the camera. The measured intensity is maximum in absence of ions while it is reduced at the position of the shadow image of an ion. Under the conditions $\cos \phi\approx 1$ and $r_{R-P_0} \gg 1$, the interference term $\propto |\text{E}_{S} \text{E}_R|$ contributes dominantly to the imaging contrast. In our experiment, $\cos \phi\approx 1$ is satisfied, and by increasing $\text{E}_R$ , the SNR can be made large enough and is eventually limited by the shot noise of the strong reference beam~\cite{supp}.

%
%
\begin{figure}[t]
\begin{center}
\includegraphics[width=\linewidth]{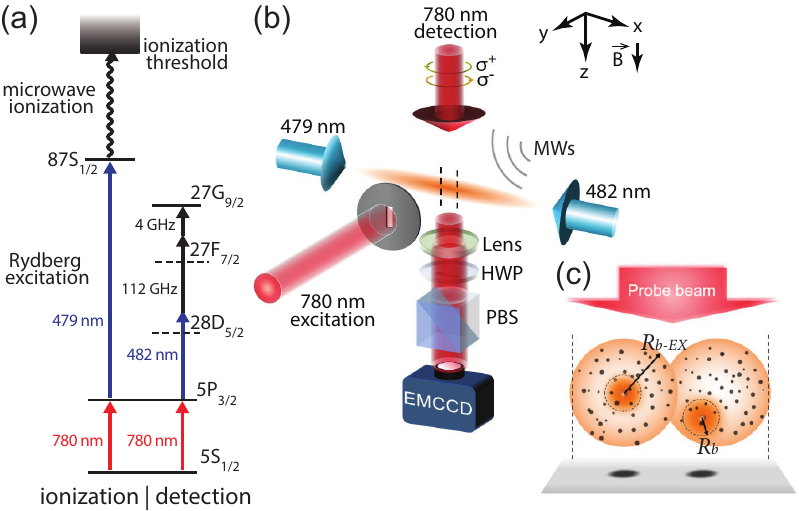}
\end{center}
\caption{\label{fig1}
(a) Energy level schemes for ion production and multi-photon EIT detection. (b) Sketch of the experimental setup. The 780 nm excitation beam reaches the atomic cloud after passing through a slit and a telescope (not shown), which limits the excitation to a segment of the atomic cloud $V_{\text{I}}$ bounded by the two dashed lines. The 780-nm beam propagating along the +$z$ direction is a superposition of the $\sigma^+$ probe field and the $\sigma^-$ reference field. After the analyzer [a half wave-plate (HWP) and a polarizing beam-splitter (PBS)], the interference image is collected by an electron-multiplying charge-coupled-device (EMCCD) camera. A bias magnetic field is applied vertically to define the quantization axis. (c) Zoom-in view at the volume $V_{\text{I}}$ exposed to excitation and ionization, with a cartoon illustration of two Rydberg ``superatoms'' (two closely packed transparent balls), ground-state atoms (black dots), and two ion-Rydberg interaction blockades (orange balls), which cast two absorption spots due to the ions on the camera screen (black spots). Here $R_b$ and $R_{b-EX}$ are defined in the text, and the dashed lines in (b) and (c) delimit the same area. Note that microwave ionization can occur anywhere within a Rydberg ``superatom'', therefore the resulting ion is not necessarily at the center of the Rydberg ``superatom''.
}
\end{figure}
%
%


Shown in Fig.~\ref{fig1} are the schematics of our experiment, and additional details can be found in the Supplemental Materials~\cite{supp}. Each experimental cycle starts with the preparation of a highly elongated, cylindrical cloud of atoms in the ground state $|g\rangle$ with a temperature of $22 \, \mu \mathrm{K}$, which has been released from an optical dipole trap for a time of flight of 10 $\mu$s. The cloud radially follows a Gaussian density distribution with a standard deviation of $\sigma_r = 5.5 \, \mu \mathrm{m}$, axially extends a few millimeters along the $x$ direction, and has a peak density of $n_0 = 7.9\times10^{11} \, \mathrm{cm}^{-3} $ that corresponds to a mean interatomic distance of $\sim$1.1 $\mu$m. To produce well-separated ions for imaging, we rely on microwave ionization of Rydberg excitations obtained in the fully blockaded regime, instead of direct photon-ionization as used previously~\cite{gross2020ion}. Rydberg excitations are induced with two lasers of wavelengths 780 nm and 479 nm, on resonance with the $|g\rangle \rightarrow |e\rangle$  and  $|e\rangle  \rightarrow |r'\rangle = |87S_{1/2}, m_J = 1/2\rangle$ transitions, respectively. While both laser pulses have a duration of 0.25 $\mu$s, the 479-nm pulse is switched on first, leading the 780-nm pulse by 30 ns. Note that only a segment of the atomic cloud of length $a_x \cong 52\ \mu$m along the $x$ direction (between the vertical dashed lines in Fig.~\ref{fig1} (b)) is exposed to the excitation lasers, while the exposure is quasi uniform radially along the $y$ and $z$ directions. This exposure volume is referred to as $V_{\text{I}}$. The excitation blockade radius $R_{b-EX} = 15.5\ \mu$m allows only one Rydberg excitation in the radial direction of the atomic cloud. More quantitative calculations suggest that driving the excitation of a high-density atomic cloud in a dissipative regime leads to unity Rydberg excitation within a blockade sphere (commonly termed as a Rydberg ``superatom''), and that two Rydberg ``superatoms'' closely packed along the axial axis are permitted within $V_{\text{I}}$~\cite{ates2012correlation,PetrosyanAdiabatic2013,PetrosyanCorrelations2013,vogt2017levy,supp}. After the laser excitation, a microwave pulse of 0.2 $\mu$s is used to ionize the Rydberg ``superatoms''~\cite{gallagher:ryd,mahon1991ionization,krug2005universal}. Rydberg excitation and microwave ionization together constitute the ion generation process, which occurs in $V_{\text{I}}$. Following the ion production, ion imaging is performed for 1~$\mu$s. The detection level scheme is the same as that in Ref.~\cite{gross2020ion} and relies on multi-photon Rydberg EIT, with an effective three-photon coupling field driving the $|e\rangle  \rightarrow |r\rangle$ transition~\cite{supp}. For homodyne detection, the intensity ratio is experimentally optimized to be $r_{R-P_0} = 18$. Note that the recorded EIT images of the elongated atomic cloud have a 1-mm-wide field of view along the $x$ direction, much larger than $a_x$.

%
%

\begin{figure}[t]
\begin{center}
\includegraphics[width=\linewidth]{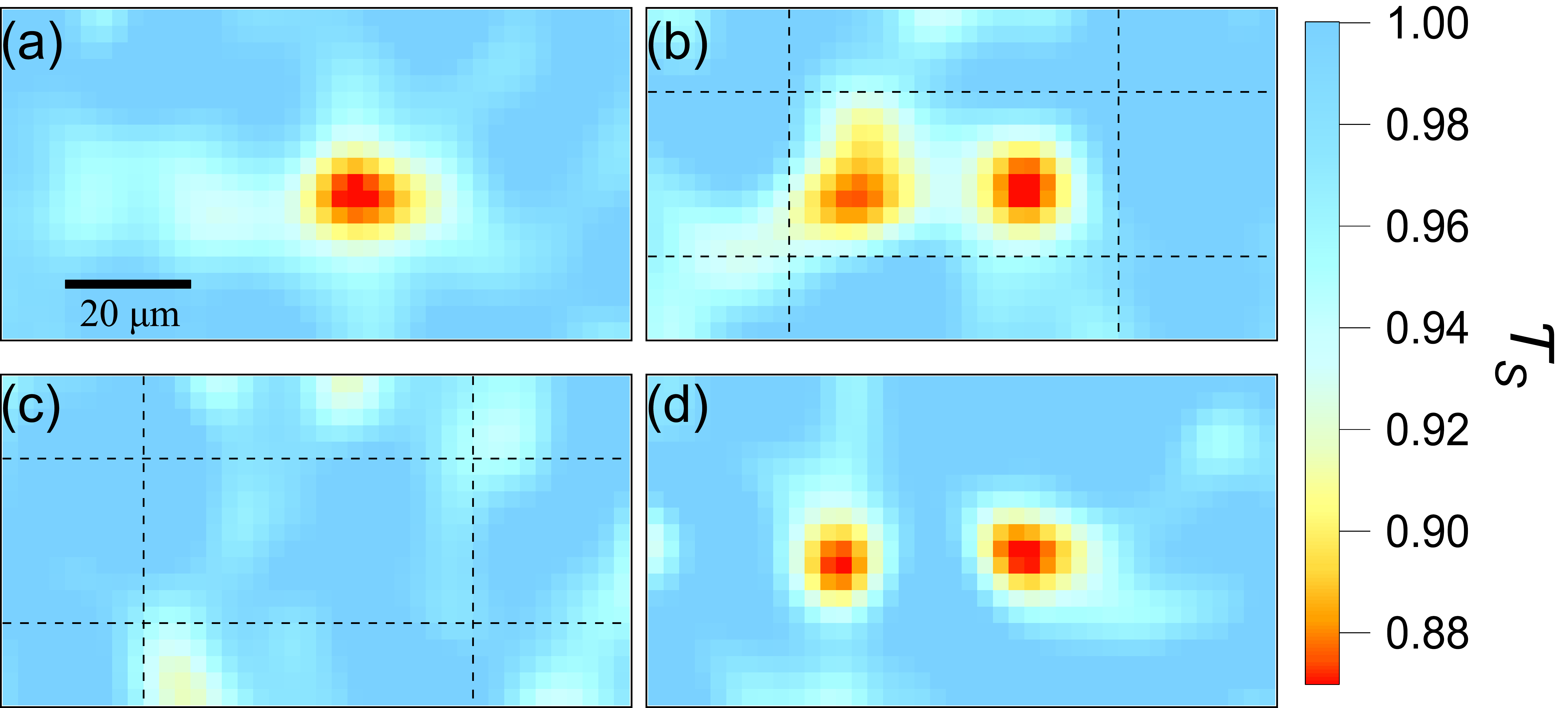}
\end{center}
\caption{\label{fig2}
Gaussian smoothed (filter size = 2 pixels) single-shot images. (a), (b) Absorption spots of one ion and two ions present in $A_{\text{I}}$, respectively. (c) No ion present in $A_{\text{NoI}}$. (d) Absorption spots of two ions from simulation. The dashed rectangles in (b) and (c) (that in (a) is not drawn) respectively indicate $A_{\text{I}}$ and $A_{\text{NoI}}$, where the horizontal dashed lines, separated by 4$\sigma_r$, correspond to the $1/e^2$ radius of the atomic cloud and the vertical ones are separated by $a_x$.
}
\end{figure}
%

In each experimental cycle, one EIT image is recorded, either with or without the ion production process. Each acquired EIT image is first processed to remove undesirable fringes~\cite{supp}. Subsequently, a single-shot EIT image recorded with preceding ion generation process is normalized at each pixel position~$(x,y)$ by a reference image to obtain a transmission distribution $\mc{T}(x,y)$. Here the reference image is generated by averaging the EIT images obtained with no ion generation and reflects the residual absorption of the atomic cloud due to the long-range interaction of state $|r\rangle$. The processed single-shot image $\mc{T}(x,y)$ may be further smoothed with Gaussian filtering into $\mc{T}_S(x,y)$. Shown in Figs.~\ref{fig2} (a) and (b) are such single-shot smoothed transmission distributions $\mc{T}_S(x,y)$ centered around the area $A_{\text{I}}$, which corresponds to the image of $V_{\text{I}}$. Pronounced absorption spots of one and two ions are clearly visible inside $A_{\text{I}}$. Meanwhile, Fig.~\ref{fig2} (c) is a different part of $\mc{T}_S(x,y)$ centered around an area $A_{\text{NoI}}$, corresponding to the image of a volume $V_{\text{NoI}}$ that is another segment of the atomic cloud identical to $V_{\text{I}}$ but not being exposed to the ion generation. No pronounced absorption spot appears in $A_{\text{NoI}}$. In the following paragraphs, we provide an unbiased method to decide whether an absorption spot pattern is due to the presence of an ion or due to imaging noises.

%
%
\begin{figure}[t]
\begin{center}
\includegraphics[width=\linewidth]{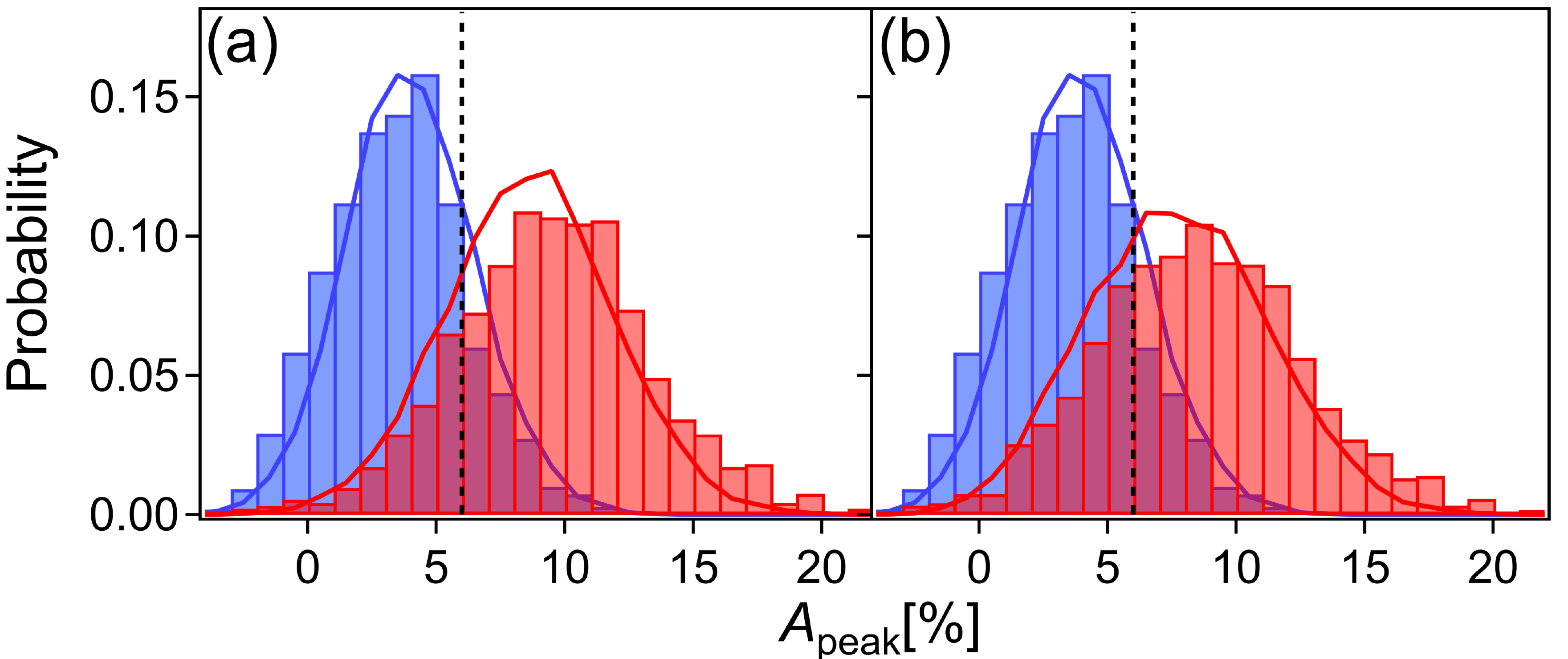}
\end{center}
\caption{\label{fig3}
Histograms of the peak absorption amplitudes of the two largest peaks (a) and of all the peaks (b) inside the area $A_{\text{I}}$ (red) vs. histogram of all the local ``absorption'' maxima inside $A_{\text{NoI}}$ (blue). The solid-line profiles are the results of theoretical simulations carried out using the experimental parameters. The vertical dashed lines indicate the threshold $A_{thld} = 6\%$. The distributions are from the statistics of about 500 single-shot images.
}
\end{figure}
%
%

To accomplish this, we construct statistical comparison of the absorption spots' amplitudes in the two imaging areas $A_{\text{I}}$ and $A_{\text{NoI}}$. In each smoothed image $\mc{T}_S(x,y)$, we find the amplitudes of local transmission minima $\mc{T}_{Smin}$ in $A_{\text{I}}$ as well as in $A_{\text{NoI}}$. In Fig.~\ref{fig3}, we plot the histograms of peak absorption amplitudes, defined as $A_{peak} = 1 -\mc{T}_{Smin}$. As stated in the discussion of Figs.~\ref{fig1} (b) and (c), there are two Rydberg ``superatoms'' in $V_{\text{I}}$, and consequently, up to two ions are expected. In Fig.~\ref{fig3} (a), we plot the histogram of only the two largest absorption peaks in $A_{\text{I}}$, mostly coming from ions, versus the histogram of all the local ``absorption'' maxima due to noises in $A_{\text{NoI}}$ (on average $\sim$2.2 per image). The two distributions are well separated, and can be best distinguished by the threshold peak amplitude $A_{thld} = 6\%$. In Fig.~\ref{fig3} (b), we plot the histogram of all the peaks in $A_{\text{I}}$ (on average $\sim$2.5 per image) versus that for $A_{\text{NoI}}$. Since there are not many additional absorption peaks in $A_{\text{I}}$ beyond the two largest ones, including them in the statistics leaves the threshold separating the two distributions almost unchanged.

We compare our experimental results with those of a model based on Maxwell-Bloch equations, which describes an ensemble of effective three-level atoms interacting with the probe and effective coupling fields in the presence of the electric field due to the ions~\cite{supp}. In addition, we consider the shot noise due to the strong reference light as well as a Gaussian noise that accounts for the effects of the camera read noise and speckle noises not fully removed by our fringe removal program. The parameters entering the simulation are obtained from experimental calibrations while the Gaussian noise level is kept as an adjustable parameter. For simulating images, each time we place a pair of ions in the atomic volume $V_{\text{I}}$ via direct sampling while taking into account the excitation blockade $R_{b-EX}$. A smoothed sample image from our simulation is displayed in Fig.~\ref{fig2} (d). We also simulate single-shot images with no ion present. The solid curves in Figs.~\ref{fig3} (a) and (b) are the distributions of peak absorption amplitudes obtained from analyzing the simulated images in the same way as the experimental ones. The good agreement between experiment and simulation confirms that we have realized single-shot imaging of individual ions and that there are mostly two ions present in the volume $V_{\text{I}}$. The overlap between the distributions with and without ions is largely due to fluctuations of peak amplitude induced by imaging noises and random ion locations along the Gaussian density profile in the radial direction of the atomic cloud. We evaluate the probability of identifying an ion in a single-shot image based on the distributions of Fig.~\ref{fig3} (a), where the true negative probability (no ion is detected when none is present) below $A_{thld}$ is about the same as the true positive probability (the presence of an ion is detected) above $A_{thld}$~\cite{Bergschneider2018}. The fidelity of detecting an ion is then defined as the true positive probability and estimated to be (80 $\pm$ 5)\%.

\begin{figure}[t]
\begin{center}
\includegraphics[width=0.95\linewidth]{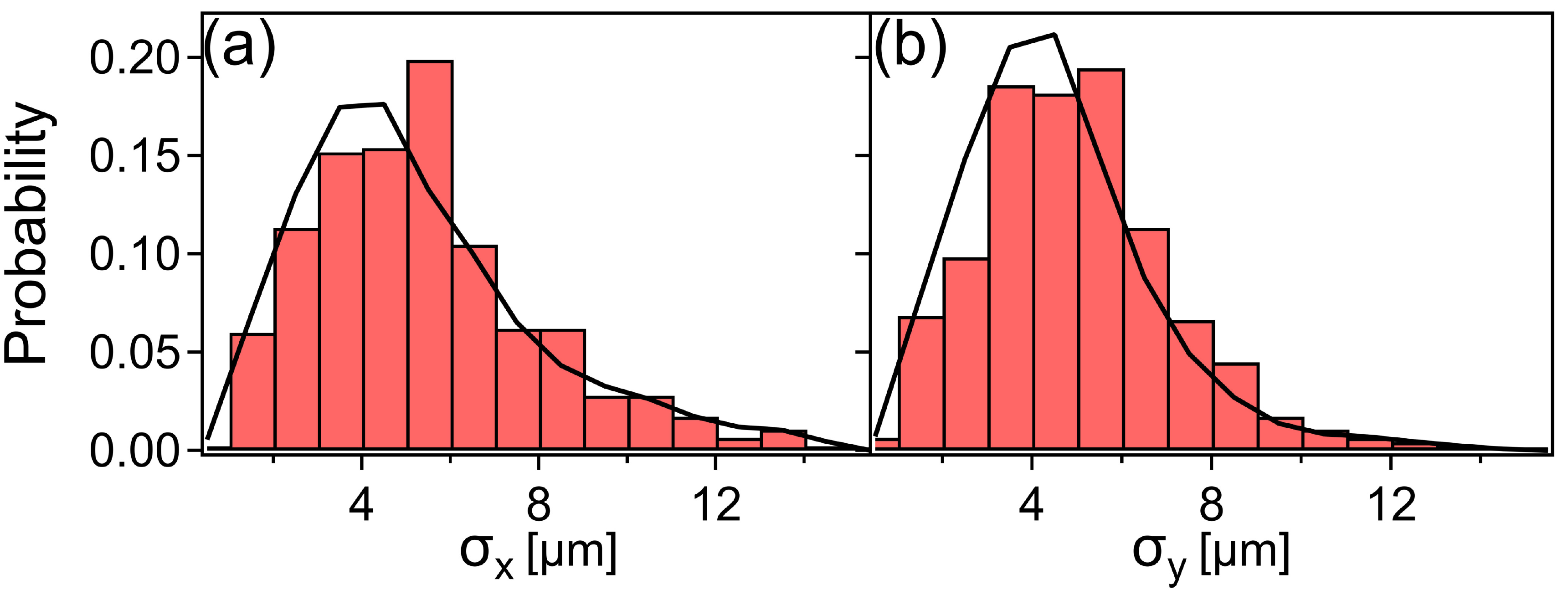}
\end{center}
\caption{\label{fig4}
(a) and (b) Histograms of the radii $\sigma_x$ and $\sigma_y$ extracted from fitting a 2-dimensional Gaussian distribution to the absorption spots of peak amplitude larger than $A_{thld}$. The solid lines are the results from theoretical simulations. The distributions are from the statistics of nearly 400 single-shot images, each of which contains up to two absorption spots of peak amplitude larger than $A_{thld}$.
}
\end{figure}

To further confirm that an absorption spot with an amplitude above $A_{thld}$ corresponds to an image of a single ion, we compare the sizes of the absorption spots in experimental images with that of simulated ones. We fit a 2-dimensional (2D) Gaussian profile to the absorption spots in $\mc{T}(x,y)$ and extract their standard deviations~\cite{supp}. The histograms of the extracted radii $\sigma_x$ and $\sigma_y$ along $x$ and $y$ directions are respectively shown in Figs.~\ref{fig4} (a) and (b), and compared to the theoretical distributions plotted as solid lines. The experimental result agrees well with that from theory, and both appear to feature spread-out distributions. This suggests that, while the sizes of single-ion absorption spots are inherently related to the ion-Rydberg interaction blockade radius $R_b= [2 C_4 / (\hbar \gamma_{EIT})]^{1/4}$ = 10.2 $\mu$m, they do vary due to imaging noises as well as the random ion locations across the Gaussian density distribution of the atomic cloud.

\begin{figure}[t]
\begin{center}
\includegraphics[width=\linewidth]{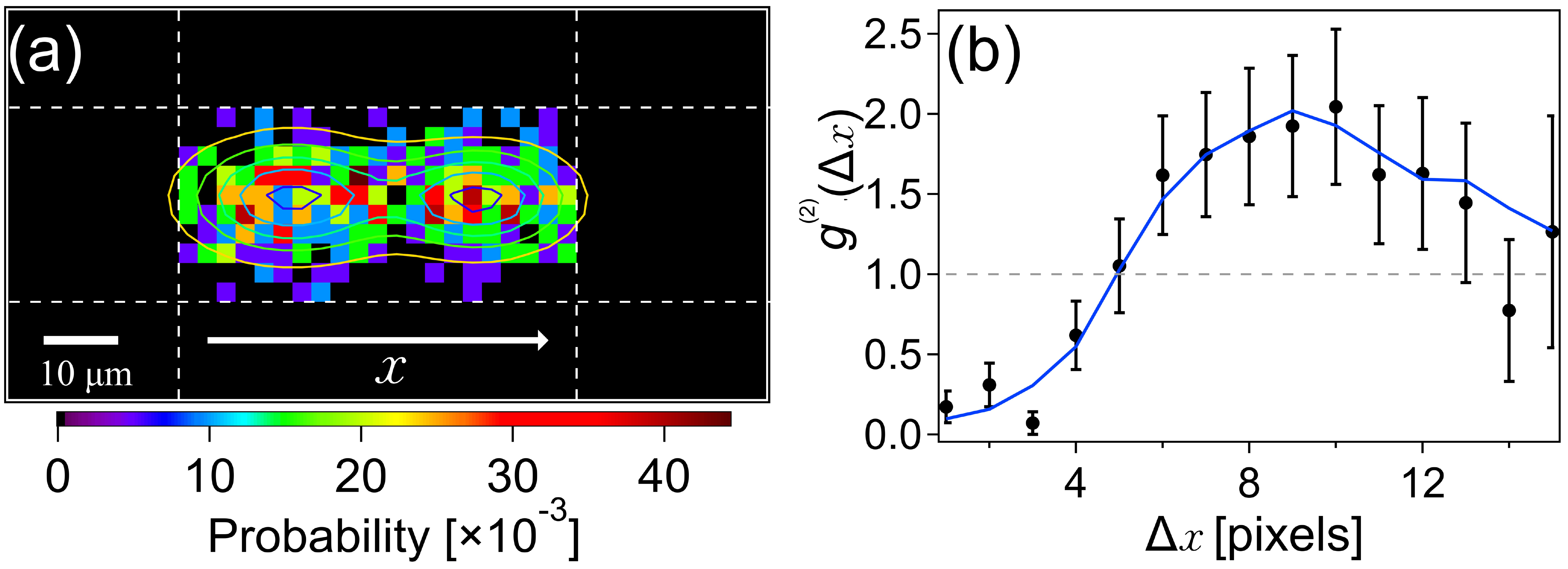}
\end{center}
\caption{\label{fig5}
(a) Two-dimensional (2D) distribution of the peak positions of two ions in a single-shot image. The dashed rectangle corresponds to the area $A_{\text{I}}$. The thin lines are the result of a 2D Gaussian contour fit. (b) Pair correlation function of the positions of two ions in single-shot images. The error bars represent the standard error over more than 200 images. For this figure, we choose the single-shot images having the two absorption spots of peak amplitude larger than $A_{thld}$.
}
\end{figure}

Finally, we use our ion imaging technique to inspect Rydberg excitations  and their spatial correlation. As discussed earlier and illustrated in Fig.~\ref{fig1}(c), our experimental conditions lead to two Rydberg ``superatoms'' in a closely packed arrangement  ~\cite{PetrosyanAdiabatic2013,PetrosyanCorrelations2013}. Each time microwave ionization is applied, a Rydberg ``superatom'' is projected into an ion located within the blockade sphere of radius $R_{b-EX}$. In Fig.~\ref{fig5} (a), we plot the 2D distribution of the positions of two ions created from the two Rydberg ``superatoms''. Here the position of an ion is taken as the location of an absorption maximum with its amplitude above the threshold $A_{peak} > A_{thld}$. Two separated distribution peaks are visible, and a 2D Gaussian fit determines the distance between the two peaks to be 10.0 $\pm$ 0.4 pixels (corresponding to 24.6 $\pm$ 1.0 $\mu$m at the position of the atoms). To further study the one-dimensional (1D) spatial correlation of the two Rydberg ``superatoms'' along the $x$ direction, we measure the 1D pair correlation function of two ions that are separated by $\Delta x$~\cite{Note1},
\begin{equation}
g^{(2)}(\Delta x) = \frac{2 \Sigma_{i} \langle n_{\mathrm{ion}}(x_i) \cdot n_{\mathrm{ion}}(x_i + \Delta x) \rangle}{\Sigma_{i} \langle n_{\mathrm{ion}}(x_i) \rangle \cdot\langle n_{\mathrm{ion}}(x_i + \Delta x) \rangle },
\end{equation}
where $n_{\mathrm{ion}}(x_i)$ (= 0 or 1) is the ion number at the pixel position $x_i$ in a single-shot image and $\langle . \rangle$ is the ensemble average over single-shot images. The pair correlation function $g^{(2)}(\Delta x)$ plotted in Fig.~\ref{fig5} (b) clearly shows anti-bunching ($g^{(2)} < 1$) at small $\Delta x$ and bunching ($g^{(2)} > 1$) at large $\Delta x$. While the anti-bunching comes from the excitation blockade, the bunching reveals that the excitation of two Rydberg superatoms is positively correlated to form an ordered arrangement. The bunching forms a broad peak around $\Delta x \sim 10$ pixels, which is consistent with the distance between the two peaks of the ion position distribution, obtained by the 2D fitting of Fig. 5(a). This separation of $\sim$ 1.6 $R_{b-EX}$ indicates that the two Rydberg ``superatoms'' are closely packed. This is generally compatible with theoretical predictions for exciting such Rydberg ``superatoms'' in a dense atomic medium~\cite{PetrosyanCorrelations2013}. Our imaging technique offers an alternative way to investigate highly-correlated Rydberg excitations~\cite{schauss2012observation}.


In conclusion, we have achieved single-shot imaging of individual ions in an ion-atom mixture by utilizing the ion-Rydberg-atom interaction blockade, and used this technique to investigate the spatial correlation of Rydberg ``superatoms''. The detection method can be further improved by increasing the transmission and reducing the noises of our optical imaging system. This direct and \textit{in situ} imaging of impurities in an atomic ensemble is well suited for studying hybrid atomic systems and for exploring impurities as probes to measure density distributions and the temperature of quantum gases~\cite{tomza2019hybridionatom,seah2019,mitchison2020}. The technique can be readily extended to imaging Rydberg excitation impurities in an atomic ensemble~\cite{gunter2013observing}, which will be an important asset in advancing Rydberg-atom platforms for quantum information technologies and for quantum simulation of many-body systems ~\cite{lukin:01,xu2021rydbergensembels}.

\section{Acknowledgement}
\begin{acknowledgments}
The authors thank Tom Gallagher, Christian Gross, and Klaus M{\o}lmer for useful discussions and acknowledge the support by the National Research Foundation, Prime Ministers Office, Singapore and the Ministry of Education, Singapore under the Research Centres of Excellence programme.
\end{acknowledgments}
%
%

%
%
%
%
\clearpage
\pagebreak
\begin{titlepage}
\begin{center}
\section*{Supplementary material: Fast Single-shot Imaging of Individual Ions via Homodyne detection of Rydberg-Blockade-Induced Absorption}
\end{center}
\end{titlepage}
\renewcommand{\thefigure}{S\arabic{figure}}
\renewcommand{\theequation}{S\arabic{equation}}
\setcounter{secnumdepth}{2}
\setcounter{equation}{0}
\setcounter{figure}{0}
\setcounter{table}{0}
\setcounter{page}{1}

\section{Generation of individual ions}
\label{Iongeneration}

Individual ions are produced by microwave ionization of Rydberg excitations in the fully blockaded regime, so that the produced ions are well-separated due to the Rydberg excitation blockade. Under the conditions of our experiment, a maximum of two Rydberg excitations are allowed in the exposure area, forming a packed arrangement due to excitation correlations. More details about the scheme are given in the following subsections.

\subsection{Rydberg excitation}

An EIT storage scheme is used for producing Rydberg excitations, where a 780-nm probe laser and a 479-nm coupling laser are on resonance with the $|g\rangle = |5S_{1/2}, F = 2, m_F = 2\rangle \rightarrow |e\rangle = |5P_{3/2}, F = 3, m_F = 3\rangle$ and $|e\rangle  \rightarrow |r'\rangle = |87S_{1/2}, m_J = 1/2\rangle$ atomic transitions, respectively (as in Fig. 1(a) of the main text). The geometry of the two excitation beams is illustrated in Fig. 1(b) of the main text. The coupling beam propagates coaxially with the atomic cloud and has its $1/e^2$ beam waist of $ 31 $ $\mu$m aligned with the center of the elongated cigar-shape atomic cloud. The coupling field is linearly polarized along the $\hat{y}$ axis. The probe beam is linearly polarized along the $\hat{x}$ axis and propagates along the $-\hat{y}$ direction, where a slit followed by a telescope in the beam path controls the size of the probe light that reaches the atoms. Only a segment of the atomic cloud of length $a_x$ along the $x$ direction (between the vertical dashed lines in Figs. 1(b) and 1(c) of the main text) is exposed to excitation, while the exposition is quasi uniform along $y$ and $z$ directions.

Our EIT excitation storage is performed according to the following time sequence. The 479-nm and 780-nm lasers are switched on and off consecutively with delays of about 30 ns. The 479-nm starts first and is switched off first for enabling the Rydberg excitation storage. The two lasers are simultaneously on for $\sim$ 220 ns. The two excitation lasers have the peak Rabi frequencies of $\Omega_{P-EX} = 2\pi \times \, 0.7$ MHz and $\Omega_{C-EX} = 2\pi \times \, 2.9$ MHz, respectively. The resulting excitation EIT linewidth $\gamma_{EIT-EX} = \Omega_{C-EX}^2/\Gamma_e = 2\pi \times \, 1.38$ MHz, where $\Gamma_e = 2\pi \times \, 6.07$ MHz is the decay rate of the intermediate state $|e\rangle$, yields an excitation blockade radius of $R_{b-EX} = (-2 C_6[|87S\rangle]/\gamma_{EIT-EX})^{1/6}$ = 15.5 $\mu$m.

The large excitation blockade radius $R_{b-EX}$ and the high peak atomic density of $n_0 = 7.9\times10^{11} \, \mathrm{cm}^{-3} $ lead to two important features: each blockade sphere contains one Rydberg excitation with nearly unity probability and the Rydberg excitation blockade spheres (or Rydberg ``superatoms'') are closely packed. According to the superatom model of strongly interacting dissipative systems, the total number of Rydberg excitation $N_R$ in a blockade sphere is~\cite{vogt2017levy,ates2012correlation,PetrosyanCorrelations2013}
\begin{equation} \label{RydExcitation}
\begin{aligned}
N_R & = N_G \times f_r \\
& = N_G \times \frac{f_0}{1-f_0+f_0 N_G},
\end{aligned}
\end{equation}
where $N_G = \int_0^{R_{b-EX}} n_{at} (r) 4\pi r^2 dr$ is the number of ground-state atoms inside the volume of the blockade sphere, $f_0 = \frac{\Omega_{P-EX}^2}{\Omega_{P-EX}^2 + \Omega_{C-EX}^2}$ is the Rydberg excitation fraction of an atom in the absence of any interaction, while $f_r$ is the Rydberg excitation fraction of an atom inside the blockade sphere. The high atomic density gives a large $N_G \sim 4200$ and a near unity Rydberg excitation per blockade sphere $N_R = 1$. Moreover, the large $N_G$ results in very strong correlation between Rydberg excitations such that their blockade spheres are packed into an ordered configuration. In our experiment, $R_{b-EX} > \sigma_r$ and $4R_{b-EX} \geq a_x$, two Rydberg ``superatoms'' appear to be side-by-side along the $x$ direction, as illustrated in Fig. 1(c) of the main text.

\subsection{Microwave ionization}
Following the excitation pulses, a microwave pulse of duration 0.2 $\mu$s is subsequently switched on to ionize the Rydberg excitations in the state $|r'\rangle = |87S_{1/2}, m_J = 1/2\rangle$. There are two thresholds for ionizing a Rydberg state with the principal quantum number $n$~\cite{gallagher:ryd}. One is the classical ionization limit $E_c = 1/(16n^4)$, and the other is the Inglis-Teller limit $E_{it} = 1/(3n^5)$, both of which are written in atomic unit of electric field. Above $E_c$, the valence electron becomes free from the ionic core and the atom ionizes rapidly. Meanwhile, at the Inglis-Teller limit, the adjacent levels of principal numbers $n$ and $n+1$ form an avoided crossing. A microwave field of strength $E_{MW} > E_{it}$ can ionize the Rydberg electron of state $n$ via the Landau-Zener effect~\cite{mahon1991ionization,krug2005universal}. For the Rydberg state $|r'\rangle$ in our experiment, the two thresholds are $E_c$ = 6.5 V/cm, and $E_{it}$ = 0.4 V/cm, respectively.

\begin{figure}[tbph]
\begin{center}
\includegraphics[width=0.9\linewidth]{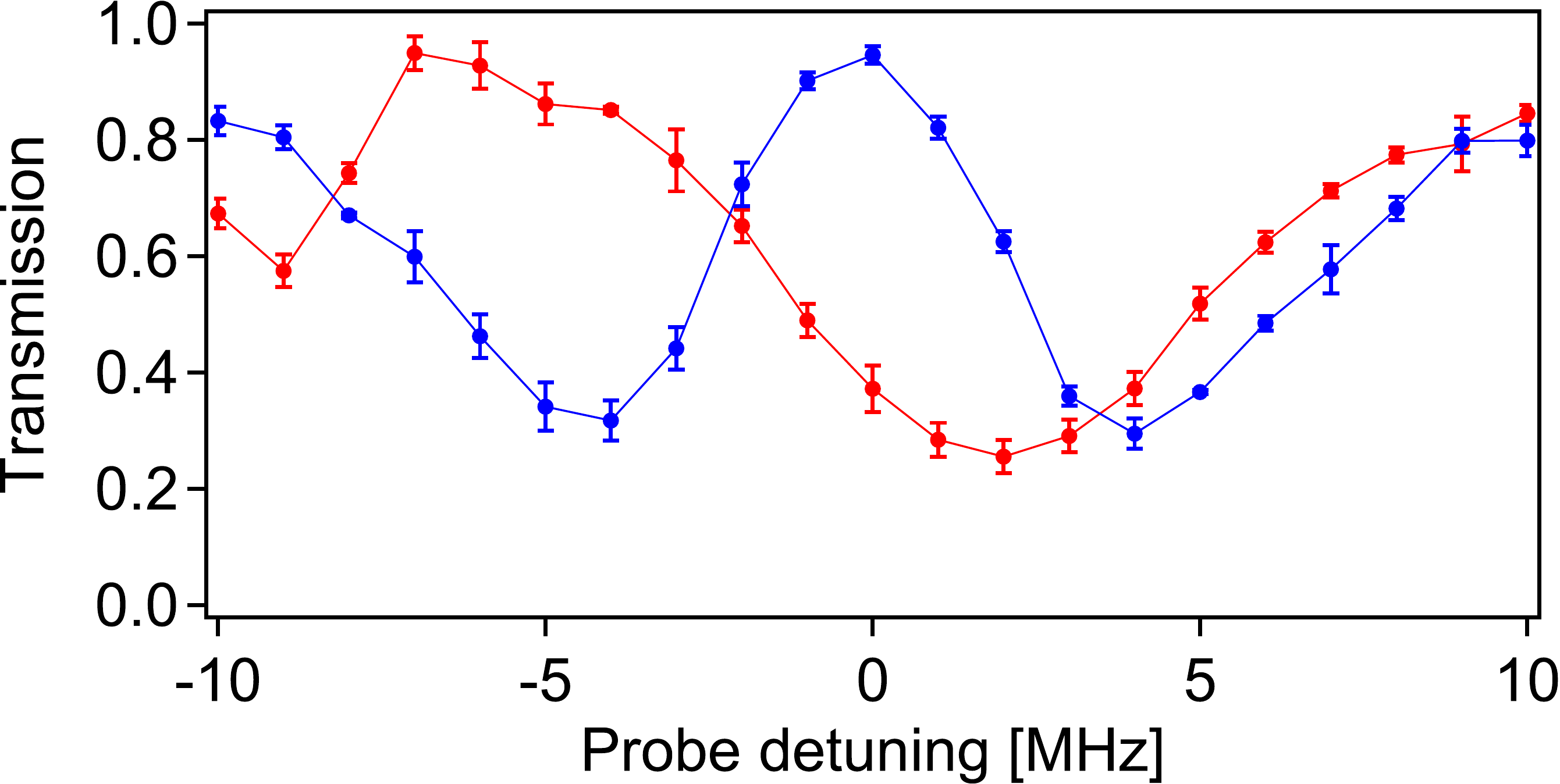}
\end{center}
\caption{\label{FigS1}
Rydberg EIT spectrum of the $|28D_{5/2}\rangle$ state, without microwave (blue) and with the presence of a microwave field of frequency $\sim$2.9 GHz (red). The error bars represent one standard deviation.
}
\end{figure}

In our apparatus, the microwave is sent into the metal vacuum chamber through a quartz window. It happens that atoms see a stronger microwave field around frequency $\nu_{MW} \sim $2.9 GHz than at other frequencies. This is likely because the metal vacuum chamber acts like a cavity to enhance the microwave coupling. The microwave field seen by the atoms can be calibrated by the shift of a Rydberg EIT resonance, with the upper level being the $|28D_{5/2}\rangle$ state. In the presence of the microwave field $E_{MW} \mathrm{cos}(2\pi \nu_{MW} t)$, this Rydberg energy level acquires an AC Stark shift, mainly due to its coupling to the nearby $|27F_{7/2}\rangle$ and $|29P_{3/2}\rangle$ states. Consider that the respective energy differences of the $|29P_{3/2}\rangle$ and $|27F_{7/2}\rangle$ from the $|28D_{5/2}\rangle$ state are $E_P $ = -104.8 GHz and $E_F$ = +112.3 GHz, and the respective dipole moments of the $|28D_{5/2}\rangle \rightarrow |29P_{3/2}\rangle$ and $|28D_{5/2}\rangle \rightarrow |27F_{7/2}\rangle$ transitions are $\mu_{DP} =$ 611 $e a_0$ and $\mu_{DF} =$ 637 $e a_0$, the shift of the $|28D_{5/2}\rangle$ state is estimated to be
\begin{equation}\label{LevelShift}
\begin{aligned}
\Delta E_{28D}  = & - (\frac{\Omega_{DP}^2}{4 (E_P- \nu_{MW}) } + \frac{\Omega_{DP}^2}{4 (E_P + \nu_{MW}) })  \\
               &- (\frac{\Omega_{DF}^2}{4 (E_F - \nu_{MW}) } + \frac{\Omega_{DF}^2}{4 (E_F + \nu_{MW}) }), \\
\end{aligned}
\end{equation}
where $\Omega_{DP(DF)} = \mu_{DP(DF)} \cdot E_{MW} / h$, $h$ denoting the Planck constant. From the measured $\Delta E_{28D} \sim$-7 MHz, as shown in Fig.~\ref{FigS1}, we can estimate the electric field strength to be $E_{MW} \sim$ 14 V/cm, larger than both the $E_c$ and $E_{it}$ thresholds for the Rydberg state $|r'\rangle = |87S_{1/2}, m_J = 1/2\rangle$ given above. With an ionization pulse duration of 0.2 $\mu$s, ionization should be very efficient. We can give a lower bound of 90\% for the ionization probability~\cite{Noel2000}.

\section{Interference imaging system}

\subsection{Parameters of imaging fields}

A multi-photon Rydberg EIT scheme is used for imaging ions in the atomic ensemble. Given in Table~\ref{detectionparameters} is a list of all the relevant parameters in regard to the probe beam and the fields D, M1, and M2 that form the effective coupling field of $\Omega_{C,\text{eff}} = 2 \pi \times$5.5 MHz, in reference to the experimental configuration shown in Fig. 1(b) of the main text.

\small
\begin{table}[tbph!]
\caption{\label{detectionparameters} Experimental parameters related to the detection fields }
			\begin{ruledtabular}
			\begin{tabular}{ll}
			  Parameters    & Value / Unit \\
                    \hline
              of the 780 nm probe beam  &    \\
			 			  \hline			
		  	 Propagation direction    & $+\hat{z}$   \\
				Polarization     & $\hat{\sigma}^+$   \\
				$1/e^2$ radius & $ 3.4 $ mm \\
				Rabi frequency $\Omega_P$ &  $2 \pi \times 0.9\; \mathrm{MHz}$  \\
                    \hline
              of the 482 nm  beam (D field) &    \\   	
                    \hline	
                Propagation direction  & $-\hat{x}$   \\
				Polarization   & $\hat{y} $   \\
				$1/e^2$ radius & $ 46.5 \, \mu$m \\
                Frequency detuning $\Delta_D$  & $ 2 \pi \times 32 \, \mathrm{MHz} $ \\
				Rabi frequency ($\sigma^+$ component) $\Omega_D$  & $ 2 \pi \times 34 \, \mathrm{MHz} $ \\
                     \hline
              of the fields M1 and M2  &    \\   	
                    \hline	
                Propagation direction  & in the $\hat{x}-\hat{y}$ plane   \\
				Polarization   & in the $\hat{x}-\hat{y}$ plane   \\
		        Frequency detuning $\Delta_{M1}$  & $2 \pi \times  \, 55 \, \mathrm{MHz} $ \\
				Frequency detuning $\Delta_{M2}$ & $- 2 \pi \times  \, 80\, \mathrm{MHz} $ \\
                Rabi frequency ($\sigma^+$ component) $\Omega_{M1}$ & $2 \pi \times  \, 42\, \mathrm{MHz} $ \\
				Rabi frequency ($\sigma^+$ component) $\Omega_{M2}$ & $2 \pi \times  \, 43\, \mathrm{MHz} $
			\end{tabular}
			\end{ruledtabular}
\end{table}
\normalsize

\subsection{Homodyne imaging}

Imaging of individual ions relies on homodyne detection in our experiment. The signal and reference fields used for realizing homodyne detection both pass through the atomic cloud. They are of $\sigma^+$ and $\sigma^-$ polarizations, respectively, and the optical frequency is adjusted for the $\sigma^+$ to be resonant with the cycling transition $|g\rangle = |5S_{1/2}, F = 2, m_F = 2\rangle \rightarrow |e\rangle = |5P_{3/2}, F = 3, m_F = 3\rangle $. Only the signal light is affected by the atomic cloud, while the $\sigma^-$ reference light hardly interacts with atoms. This is because the $\sigma^-$ has a detuning of $2\pi\times23$~MHz from its nearest allowed transition in the presence of the 12.8 G bias magnetic field, and this nearest allowed transition is considerably weaker than the transition driven by the $\sigma^+$ light.

The ratio between $\sigma^+$ and $\sigma^-$ lights is controlled via a variable wave retarder from Meadowlark (model LRC-300-IR1). In the following discussion, we define as $\vec u_x$ and $\vec u_y$ the two unit vectors corresponding to the neutral axes of the wave retarder. A linearly polarized field with polarization axis making a $\pi/4$ angle with $\vec u_x$ and $\vec u_y$ is sent onto the wave retarder.

%
\begin{figure}[t!]
\begin{center}
\includegraphics[width=0.8\linewidth]{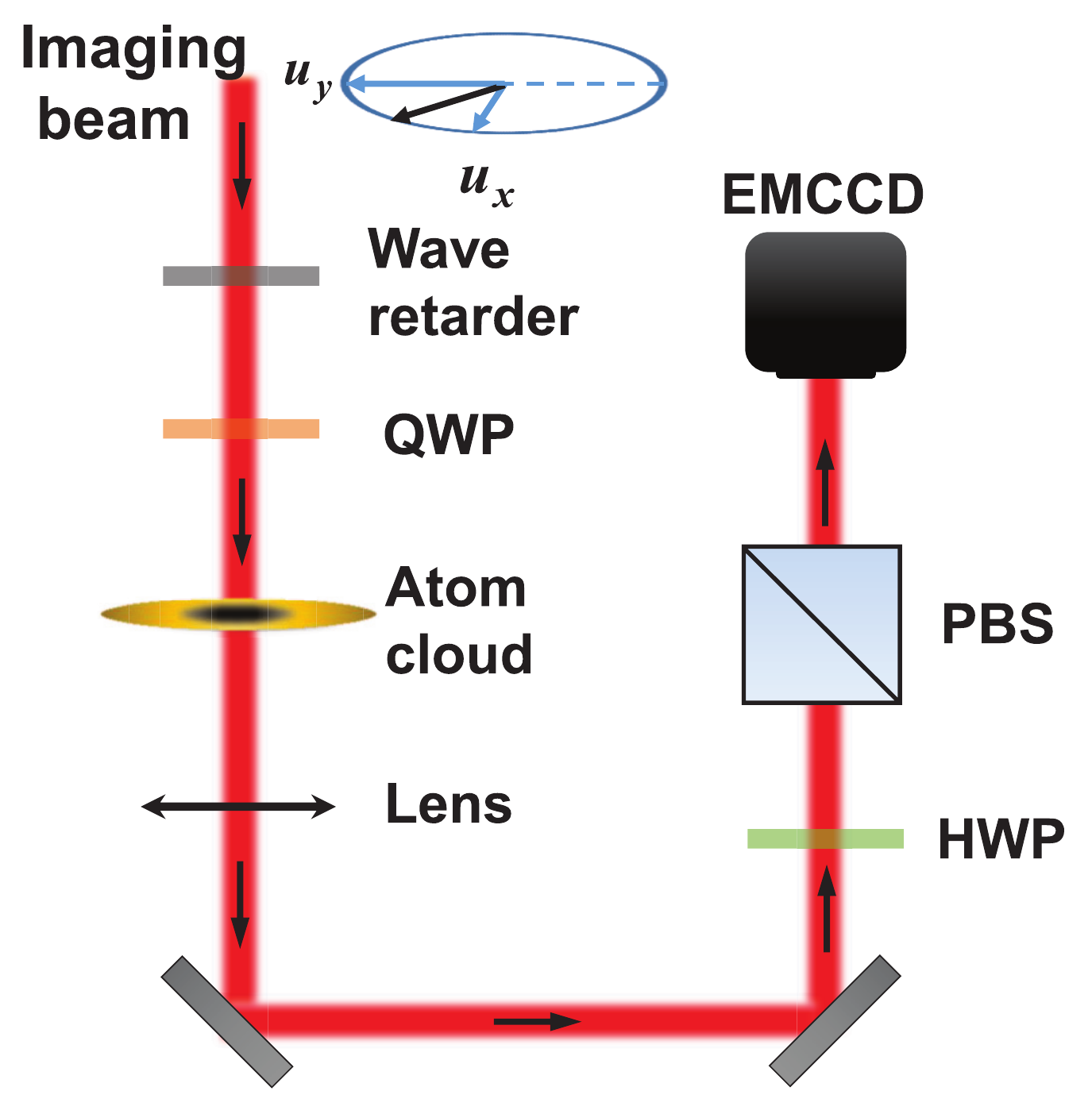}
\end{center}
\caption{\label{FigS2}
Homodyne imaging setup. A linearly polarized beam passes through a wave retarder and a quarter wave plate (QWP) whose axes are positionned at $45^{\circ}$ from that of the incident field. Next the beam crosses the atomic cloud. After transmission through the optical imaging system, the two polarization components of the beam are being interfered with a half wave plate (HWP) and a polarizing cube beamsplitter (PBS).
}
\end{figure}
%

After passing through the wave retarder, the complex field amplitude is given as

\begin{eqnarray}\label{initialfield}
\vec E_{\mathrm{in}}
= \frac{E_0}{2\sqrt 2}
\begin{pmatrix}
1\\ e^{i \varphi}
\end{pmatrix},
\end{eqnarray}
where $\varphi$ is the differential phase shift induced by the wave retarder and $E_0$ is the amplitude of the field before the wave retarder. In the experiment as shown in Fig.~\ref{FigS2}, a quarter wave plate (QWP) is added after the wave retarder for technical convenience. Its neutral axes match that of the wave retarder such that $\varphi + \pi/2$ is the total differential phase introduced by the wave retarder and the QWP together. The field $\vec E_{\mathrm{in}}'$ after the QWP is obtained by replacing $\varphi$ in Eq.~(\ref{initialfield}) with $\varphi + \pi/2$ and can be simply cast as

\begin{eqnarray}\label{incidentfield}
\vec E_{\mathrm{in}}'
=& \frac{E_0 e^{i \varphi/2}}{2 \sqrt 2}\left[\cos \left(\frac{\varphi}{2}\right)
\begin{pmatrix}
1\\ i
\end{pmatrix}-i\sin \left(\frac{\varphi}{2}\right)
\begin{pmatrix}
1\\ -i
\end{pmatrix}\right]\notag \\
=&-\frac{E_0 e^{i \varphi/2}}{2}\left[\cos \left(\frac{\varphi}{2}\right)
\vec e_+ +i\sin \left(\frac{\varphi}{2}\right) \vec e_-\right],
\end{eqnarray}
where $\vec e_{\pm}=\frac{\mp \left(\vec u_x\pm i\vec u_y \right)}{\sqrt{2}}$ are the spherical basis vectors corresponding to the $\sigma^{\pm}$ field components. Controlling $\varphi$ with the variable wave retarder allows for controlling the intensity ratio $r=\tan^2 \left(\frac{\varphi}{2}\right) $ between the two light components.

After passing through the atomic cloud, the $\sigma_+$ component acquires a phase shift $\phi$ and is transmitted with transmission coefficient $T$, whereas the $\sigma_-$ component is unaffected in first approximation.
Hence the total field writes
\begin{eqnarray}
\vec E_{\mathrm{at}} \propto \sqrt{T} \exp \left(i \phi\right) \cos \left(\frac{\varphi}{2}\right)
\vec e_+ +i\sin \left(\frac{\varphi}{2}\right) \vec e_-.
\end{eqnarray}
In a three-level ladder system, we have got $\phi=0$ when the $\sigma^+$ probe field is resonant with the corresponding atomic transition. This remains true in our effective three-level system  where imaging is performed in conditions of EIT resonance. Assuming that the $\sigma^-$ component is unaffected is justified as the absorption of the $\sigma^-$ light is less than 1\% and the acquired phase does not exceed $1^{\circ}$ given the frequency shift of $2 \pi \times 23$~MHz between the resonances of the allowed $\sigma^+$ and $\sigma^-$ transitions at the bias field of 12.8 G used in our experiment.

Then the two fields are sent onto an analyzer composed of a rotating half wave plate (HWP) and polarizing cube beamsplitter (PBS) for interfering both field components. It is assumed that the polarization axes of the PBS match that $(\vec u_x, \vec u_y)$ of the wave retarder. Considering the rotation angle $\beta$ of the HWP's neutral axes with respect to $(\vec u_x, \vec u_y)$, the fields components along $\vec e_{\pm}$ are transformed as $\vec e_{\pm}\rightarrow - e^{\mp 2 i \beta} \vec e_{\mp}$. Consequently, the total field before the PBS is given by
\begin{eqnarray}
\vec E_{\mathrm{PBS}} \propto \sqrt{T} e^{i \phi} e^{-2 i \beta}  \cos \left(\frac{\varphi}{2}\right)
\vec e_- +i e^{2 i \beta} \sin \left(\frac{\varphi}{2}\right) \vec e_+.\notag
\end{eqnarray}
The camera measures the intensity along $\vec u_x$ after the PBS.
The recorded intensity is of the form
\begin{align}
I \propto  T \cos^2 \left(\frac{\varphi}{2}\right)&
+ \sin^2 \left(\frac{\varphi}{2}\right)\notag \\
& + \sqrt{T}\sin \left(\varphi\right)\sin \left( 4 \beta - \phi\right),\label{IntensityExpres}
\end{align}
where the proportionality factor is equal to $\frac{\epsilon_0 c |E_0|^2 T_i}{4 M^2}$ with $\epsilon_0$ the vacuum permittivity, $c$ the velocity of light, and $T_i = 0.72$ and $M = 6.5$ being the transmission and the magnification of the imaging lens system, respectively. For $\phi\approx 0$, the interference signal is maximum when $\beta=\pi/8$. In practice, in the experiment we vary the ratio $r$ of intensities while keeping the intensity of the $\sigma^+$ light $\propto |E_0|^2  \cos^2 \left(\frac{\varphi}{2}\right)$ constant.
Assuming $\beta=\pi/8$, we may recast Eq.~\eqref{IntensityExpres} as follows
\begin{align}
I=  \frac{\epsilon_0 c T_i |E_0|^2}{4 M^2} \cos^2 \left(\frac{\varphi}{2}\right) \left[T + r + 2 \sqrt{r} \sqrt{T}\cos\phi \right].\label{IntensityFinal}
\end{align}
We may assume that at the position of one ion $T \approx 0$ if the optical depth due to one ion is large, whereas $T = 1$ away from this position. The interference term $\propto 2 \sqrt{r} \sqrt{T}\cos\phi $ provides the imaging contrast when the conditions $\cos \phi\approx 1$ and $r \gg 1$ are satisfied. Given the read noise count per pixel $N_{rn}$, the minimum required reference field $\text{E}_{R}$ is obtained from equating signal and noise counts
\begin{equation}
\frac{\epsilon_0 c T_i |E_0|^2 \cos^2 \left(\frac{\varphi}{2}\right) \sqrt{r} }{2 h \nu M^2}  A \delta t =  N_{rn},
\end{equation}
where $A$ = 256 $\mu$m$^2$ is the pixel area of EMCCD, $\nu$ is the probe photon frequency, and $\delta t$ = 1 $\mu$s is the imaging pulse duration. In theory, the signal-to-noise saturates for large $r$ due to shot noise of the reference field. Moreover, unwanted technical noise due to residual fringes is also detrimental at large $r$. In our experiment, an intensity ratio of the value $r = 18$ gives an overall best resolution in imaging individual ions in a single shot, as shown below with the results and analysis.

\section{Data analysis}

\subsection{Single-shot image of individual ions}

%
\begin{figure}[t]
\begin{center}
\includegraphics[width=0.9\linewidth]{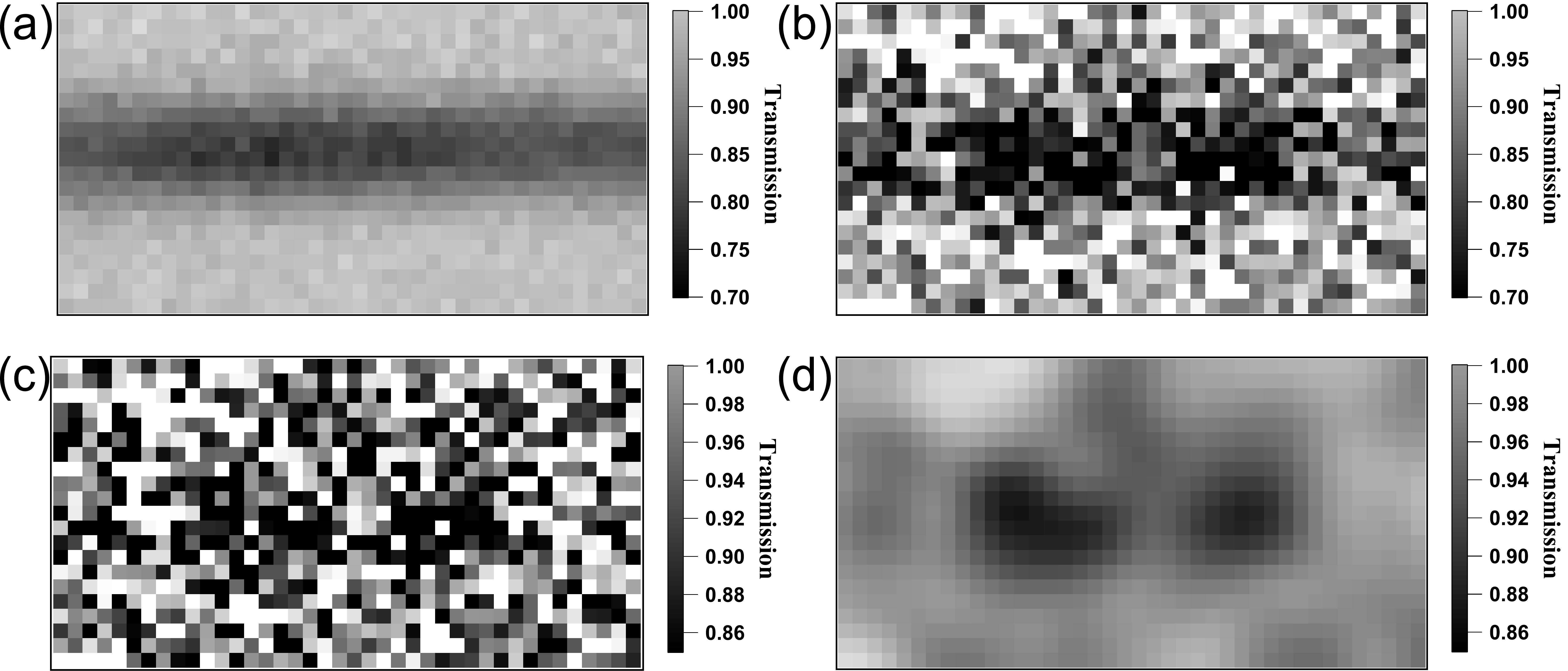}
\end{center}
\caption{\label{FigS3}
Image processing. (a) Reference EIT image. (b) Single-shot EIT image acquired with the presence of ions. (c) Processed single-shot image of ions $\mc{T}(x,y)$, obtained by dividing (b) by (a). (d) Single-shot image $\mc{T}_S(x,y)$ after smoothing (c) with a Gaussian Fourier filtering of size = 2 pixels.
}
\end{figure}
%

As described in the main text, the acquired EIT image of each experimental cycle, regardless whether ions are generated or not, is first pre-processed to remove undesirable fringes with an advanced algorithm~\cite{song2020fringeremoval}. Shown in Fig.~\ref{FigS3}(b) is such a single-shot image after fringe removal, and this particular one is taken with ions present. After fringe removal, single-shot EIT images without ions are then averaged to generate a reference image, which reflects the residual absorption of the atomic cloud due to the long-range interaction of state $|r\rangle$~\cite{gross2020ion}. Shown in Fig.~\ref{FigS3}(a) is the reference EIT image, averaged over 300 single-shot images acquired without the ion generation process. The single-shot image in Fig.~\ref{FigS3}(b) is then divided by the reference image in Fig.~\ref{FigS3}(a) to obtain the processed single-shot image $\mc{T}(x,y)$ shown in Fig.~\ref{FigS3}(c). As the normalization eliminates the residual absorption, $\mc{T}(x,y)$ highlights the increased absorption due to the presence of ions and is the starting point for further quantitative analyses. Finally, the transmission distribution $\mc{T}(x,y)$ in Fig.~\ref{FigS3}(c) is smoothed with Gaussian filtering into $\mc{T}_S(x,y)$ in Fig.~\ref{FigS3}(d), where the enhanced signal-to-noise ratio after smoothing renders the absorption spots due to ions clearly visible. The images of Figs. 2(a)-(c) of the main text are obtained in the same way. The field of view of the images in Figs.~\ref{FigS3} is the same as that of the images in Figs. 2(a) and 2(b) of the main text, i.e. centered around the area $A_{\text{I}}$, which corresponds to the image of $V_{\text{I}}$.

\subsection{Size of ion absorption spots}

\begin{figure}[!tbph]
\begin{center}
\includegraphics[width=0.8\linewidth]{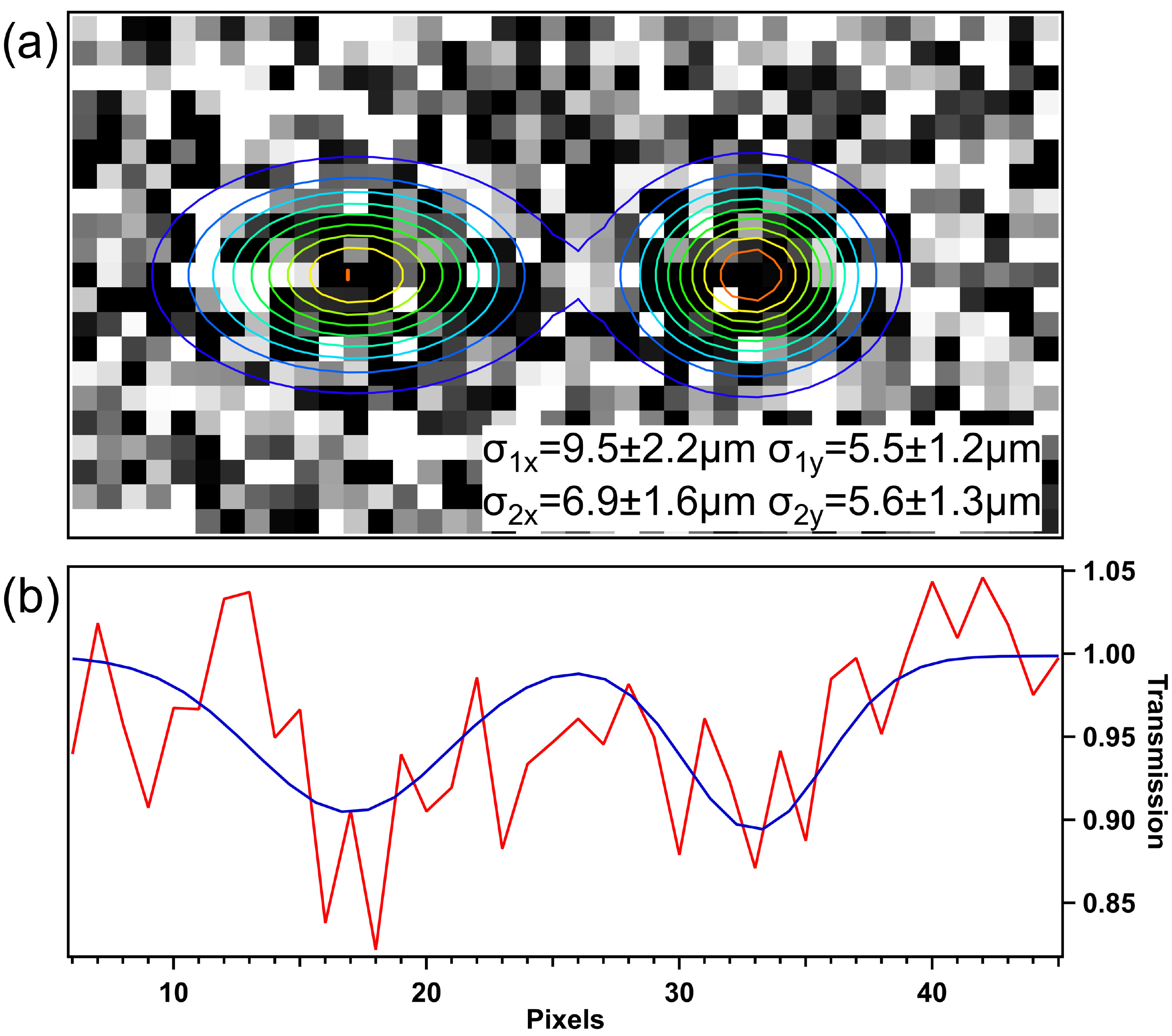}
\end{center}
\caption{\label{FigS4}
(a) Two-dimensional (2D) Gaussian fit to Fig.~\ref{FigS3}(c) for extracting the size of the shadow spots due to the absorption by the ions. (b) One-dimensional (1D) profiles obtained by integration of the 2D experimental image (red) and of the 2D fitting (blue) in Fig.~\ref{FigS4}(a) along the y (vertical) direction.
}
\end{figure}

To extract the size of ion absorption spots, a non-linear least square fit of a 2D gaussian model to an unsmoothed single-shot image is performed, using the lmfit python package~\cite{newville2018}. Here we first find the peak positions of the absorption peaks of amplitude above the threshold ($A_{peak} > A_{thld}$) from smoothed images $\mc{T}_S(x,y)$, then fix these peak positions in the 2D fitting to the unsmoothed images $\mc{T}(x,y)$. We use unsmoothed images to avoid the artificial broadening from Gaussian filtering. The fitting result is illustrated in Fig.~\ref{FigS4}, and the fitted values of the spot sizes along the $y$ and $x$ directions are given in the figure as well. Note that theoretically simulated images are fitted in the same way. The statistics of the fitting results of both experimental and theoretical images are given in Fig. 3 of the main text.

%

\subsection{Number of ions per image}

\begin{figure}[!h]
\begin{center}
\includegraphics[width=0.7\linewidth]{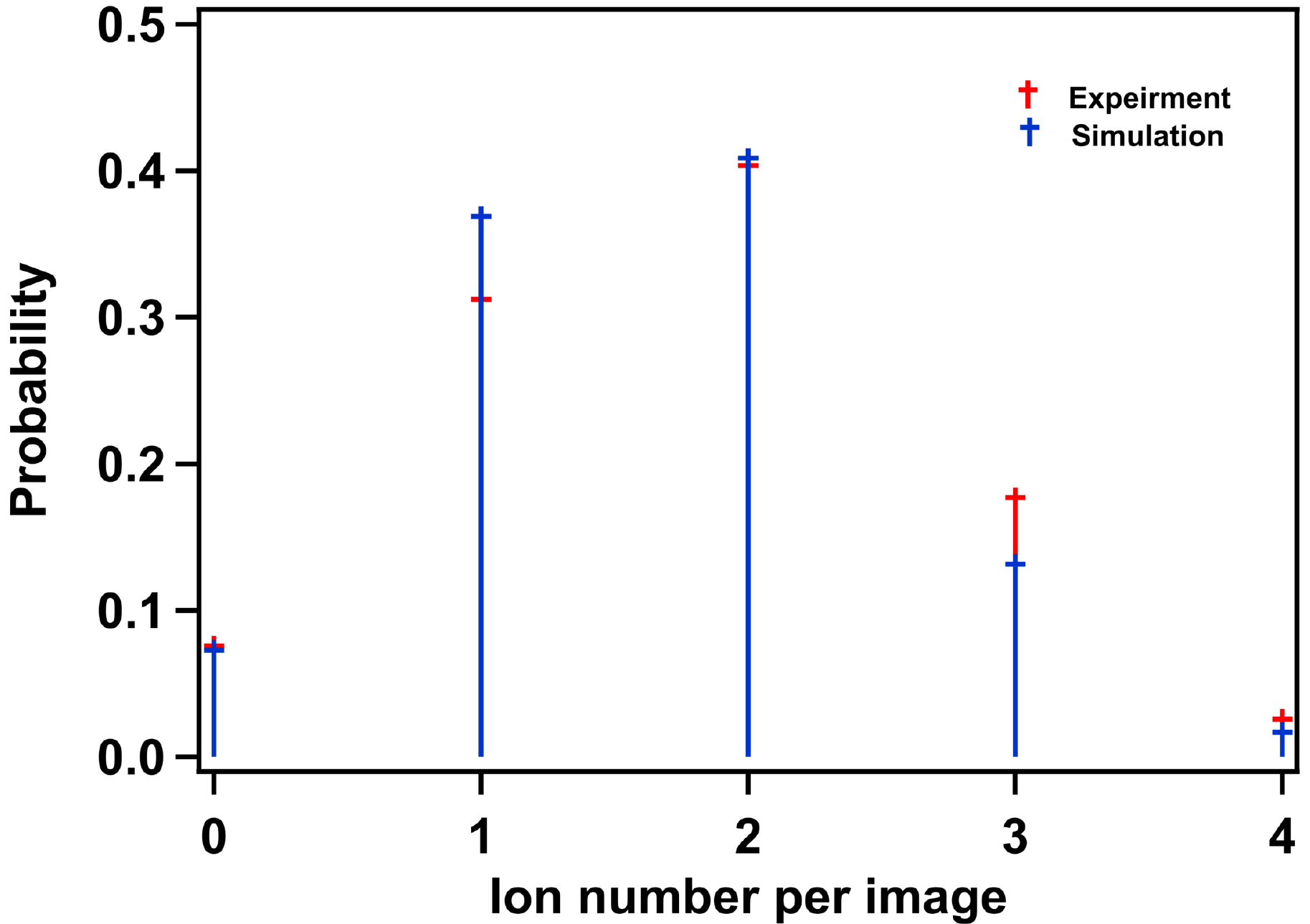}
\end{center}
\caption{\label{FigS5}
Probability distributions of the detected ion number per image both in experiment (red) and in simulation (blue).
}
\end{figure}
%

As stated in the discussion of Fig. 3 of the main text, the broad distribution in the absorption amplitudes and the consequent overlap between two distributions are ``largely due to fluctuations of peak amplitude induced by imaging noises and random ion locations along the Gaussian density profile in the radial direction of the atomic cloud.'' This also causes the variation in the number of ions detected per image. Shown in Fig.~\ref{FigS5} are the probability distributions of the detected ion number per image (the number of peaks above the absorption threshold $A_{thld} = 6\%$), from both experimental and simulated single-shot images. In the simulation, we input two ions/superatoms per image, but fluctuations of peak amplitude result in the identified ion number per image to deviate from two. In the experiment, we do not have an independent way to calibrate the number of superatoms/ions. Estimated from the experimental parameters, the probability of exciting two superatoms is nearly unity and the probability of ionizing a superatom has a lower bound of 90\%. The relevant discussions are given in Section~\ref{Iongeneration} of this Supplementary Material. The observed distributions of the detected ion number per image from experiment and simulation largely agree with each other, as shown in Fig.~\ref{FigS5}.

\subsection{One-dimensional profile of Rydberg ``superatoms''}

\begin{figure}[!h]
\begin{center}
\includegraphics[width=0.7\linewidth]{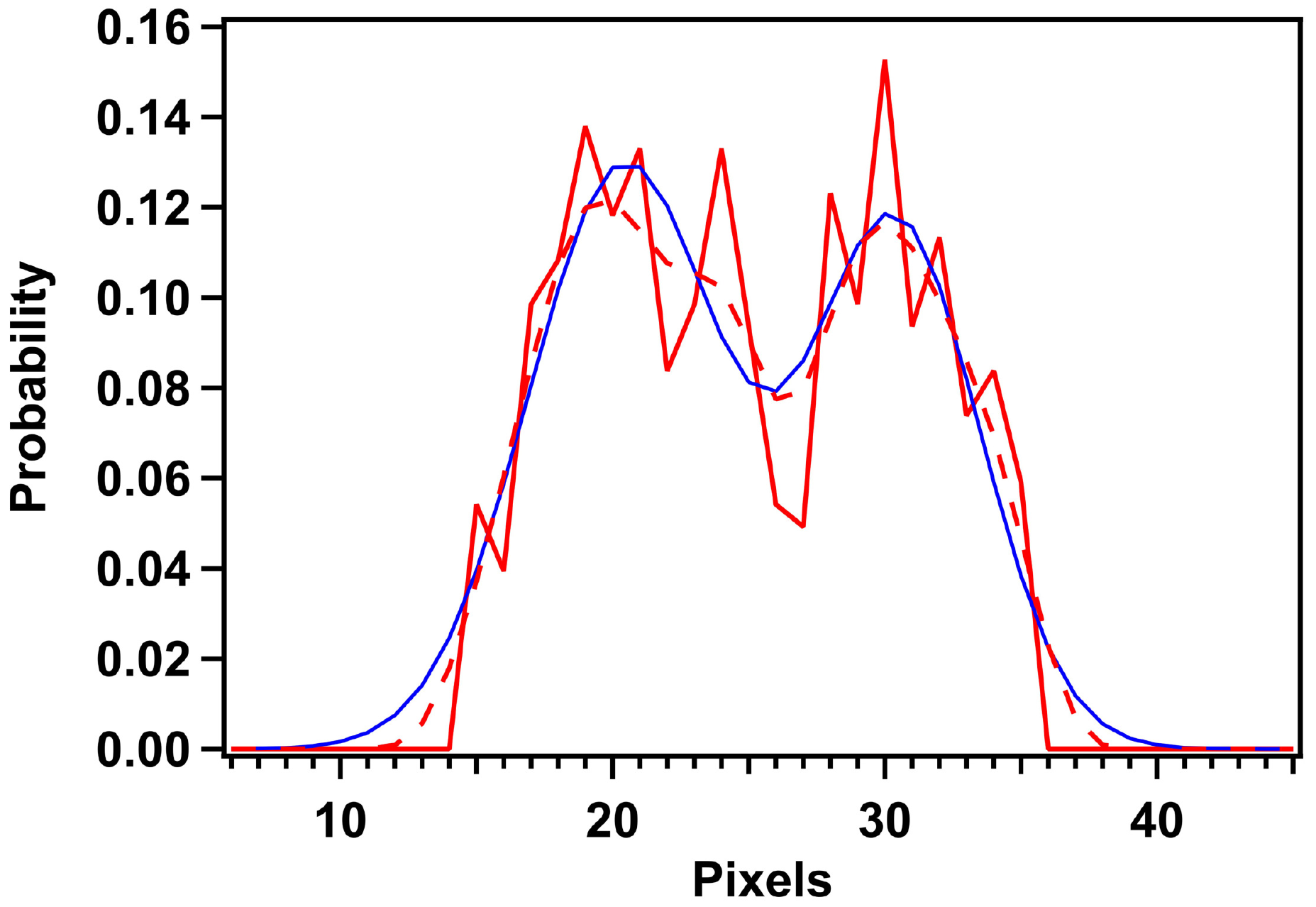}
\end{center}
\caption{\label{FigS6}
One-dimensional profiles from integrating along the vertical axis (the y-axis) both the 2D experimental distribution (red solid-line) and the 2D Gaussian fitting (blue solid-line) in Fig. 5(a) of the main text. The red dashed-line in the figure is from smoothing the red solid-line profile.
}
\end{figure}
%

Shown in Fig.~\ref{FigS6} are the 1D distributions obtained from integrating along the vertical axis (the y-axis) both the 2D experimental distribution and the 2D Gaussian fitting in Fig. 5(a) of the main text. A one-pixel local maximum around the center of the red solid curve is most likely due to noise and can be smoothed out, as shown by the dashed line in red.

\section{Theoretical simulation}

We follow the physical principle of Ref.~\cite{gross2020ion} in simulating images of individual ions. Rydberg excitation is simulated by direct sampling of pairs of hard balls of radius $R_{b-EX}/2$. Specifically, we only accept sampling of two locations of excitation in the atomic cloud when the distance between the two is larger than $R_{b-EX}$, as assumed from the hard ball model~\cite{vogt2017levy,krauth2006statistical,ates2012correlation,PetrosyanCorrelations2013}. We neglect the motion of the ions during imaging as the inter-ion distance for two ions initially separated by about 1.6$R_{b-EX}$ (the most probable separation) does not change by more than one pixel (2.46 $\mu$m) during the 1$\mu$s detection time. We then calculate the propagation of the probe field $\mc{E}_P$ through the atomic cloud in steady-state using the differential equation $\partial_{z}\mc{E}_P(\vec r)=  i  \pi /\lambda_P \; \chi(\vec r) \mc{E}_P(\vec r)$, where $\chi(\vec r)$ is the linear susceptibility at position $\vec r$. For our effective three-level system, $\chi(\vec r)$ is approximated as ~\cite{vogt2018microwave,gross2020ion}
\small
\begin{align}
\chi \left ( \vec r \right)&= \frac{i\, n_{at} \left( \vec r \right) \Gamma \sigma \lambda_P/ 4\pi}{\frac{\Gamma}{2}-i (\Delta _{P} - \delta_{1})+\frac{\Omega _{C,eff}^2 /4}{ \gamma -i \left[\Delta _P + \Delta_{C} - \delta_2 - \delta_{Stark}\left( \vec r \right)\right]}}
\label{susceptibility},
\end{align}
\normalsize
where $\Gamma/ 2 \pi= 6.067\; \mathrm{MHz}$ is the spontaneous decay rate from state $|e \rangle$, $n_{at}\left( \vec r \right)$ is the atomic density, $\gamma$ is a dephasing rate of the atomic coherence between the $|g\rangle$  and $|r\rangle$ states, and $\sigma$ is a scattering cross-section. Moreover, $\Delta_{C}$ is the detuning of the effective coupling field from the $|e \rangle \rightarrow |r\rangle$ atomic transition, and $\delta_{1}$ and $\delta_{2}$ are the AC Stark shifts affecting the $|e \rangle$ and $|r\rangle$ energy levels, respectively. Finally, $\delta_{Stark}\left( \vec r \right)=\alpha(\theta) \, \mathbf{E}\left( \vec r \right)^2 / 2$ is the Stark shift of $|r\rangle$, where $\alpha(\theta)$ is the polarizability depending on $\theta$, the angle between the quantization axis and the direction of the electrostatic field $\mathbf{E}$ of the ion at the position $\vec r$. All the other parameters in Eq.~\ref{susceptibility} are experimentally calibrated. The interference images are calculated in all the simulations following Eq.~\ref{IntensityFinal}. For the on-resonant probe field in our detection scheme, the phase shift $\phi$ is taken into account but is found to be small and the coefficient $\cos(\phi)$ remains greater than 0.95 in the vicinity of an ion. Moreover, we add in our simulation the Poisonnian Shot noise and Gaussian noise. The latter accounts for the experimentally calibrated camera read noise as well as noises due to our optical system that are imperfectly removed by the fringe removal program. A simulated image after smoothing is shown in Fig. 2(d) of the main text.
\end{document}